\documentclass[useAMS,usenatbib]{mn2e_mod}
\usepackage{graphicx}
\usepackage{amsmath}
\usepackage{amssymb}
\usepackage{deluxetable}

\def\nad{\mbox{Na\hspace{.5pt}{I}\,D}}
\def\nadone{\mbox{Na\hspace{.5pt}{I}\,D$_1$}}
\def\nadtwo{\mbox{Na\hspace{.5pt}{I}\,D$_2$}}
\def\h1{\mbox{H\hspace{.5pt}{I}}}
\def\k1{\mbox{K\hspace{.5pt}{I}}}
\def\lya{Ly$\alpha$}

\newcommand\ebv{${E_{\rm B-V}}$\relax}

\title[Sodium \& Dust]{An Empirical Relation between Sodium Absorption and Dust Extinction}

\author[Poznanski et al.]
{Dovi Poznanski$^{1}$\thanks{dovi@astro.tau.ac.il},
J. Xavier Prochaska$^{2}$,
and Joshua S. Bloom$^{3,4}$\\
$^{1}$School of Physics and Astronomy, Tel-Aviv University, Tel Aviv 69978, Israel.\\
$^{2}$Department of Astronomy and Astrophysics \& UCO/Lick Observatory, University of California, Santa Cruz, CA 95064, USA.\\
$^{3}$Department of Astronomy, University of California,
  Berkeley, CA 94720-3411, USA.\\
$^{4}$ Lawrence Berkeley National Laboratory, 1 Cyclotron Road, Berkeley, CA 94720, USA.\\
}
\begin{document}
	
\maketitle

\label{firstpage}
\begin{abstract}
Dust extinction and reddening are ubiquitous in astronomical observations and are often a major source of systematic uncertainty. We present here a study of the correlation between extinction in the Milky Way and the equivalent width of the \nad\ absorption doublet. Our sample includes more than 100 high-resolution spectra from the Keck telescopes and nearly a million low-resolution spectra from the Sloan Digital Sky Survey (SDSS). We measure the correlation to unprecedented precision, constrain its shape, and derive an empirical relation between these quantities with a dispersion of order 0.15 magnitude in \ebv. From the shape of the curve of growth we further show that a typical sight line through the Galaxy, as seen within the SDSS footprint, crosses about three dust clouds. We provide a brief guide on how to best estimate extinction to extragalactic sources such as supernovae, using the \nad\ absorption feature, under a variety of circumstances.

\vspace{1cm}
\end{abstract}

\vspace{1cm}
\section{Introduction}\label{s:intro}

%


One cannot avoid observing through the dust that permeates galaxies. Dust both reddens and absorbs radiation in ways that are usually difficult to quantify. These effects depend on the amount of dust along the line of sight as well as on the dust properties such as chemical composition and grain size distribution (see \citealt{draine03} for a review). Effectively, every observation (most notably in the UV and optical regimes) suffers from a systematic and usually unknown effect. 

A tool often-used to correct for the extinction and reddening incurred 
is the strength of the absorption doublet of sodium \nad\  $\lambda\lambda5890,5896$, a well-known tracer of gas, metals, and dust. The doublet absorption strength is generally expected to correlate with the amount of dust along the line of sight. \citet{richmond94} have shown, using 57 high-resolution stellar spectra taken by \citet{sembach93}, that the equivalent widths (EWs) of the individual components of \nad\ do indeed correlate with the color excesses measured for these stars, albeit with a noticeable scatter. \citet{munari97} add a body of 32 stars observed with somewhat lower resolution ($R \approx$ 16,500 vs. $R \approx$ 60,000--600,000); they constrain the nonlinearity of the relation, following the curve of growth, and discuss its pitfalls. They find that the precision is limited to about $\delta$\ebv$\approx 0.05$--0.15\,mag for a given measurement. They also show how at $\mathrm{EW}>0.5$\,\AA\ the \nadtwo\ line saturates and the relation flattens\footnote{We note that MZ97 invert the notations for the two lines. We will follow the traditional Fraunhofer notation, assigning D$_1$ to the redder line at $\lambda$5897 and D$_2$ to the bluer, typically stronger, line at $\lambda$5891.}. At low extinctions, and therefore low column densities, the EW is supposed to track \ebv\ linearly.

This useful correlation in high resolution spectra was recently shown not to extend in any practical way to low resolution spectra \citep[][see also \citealt{blondin09}]{poznanski11}. While the correlation survives the degradation in resolution, the large scatter in the relation is combined with confusion due to the blending of the doublet. In addition, there can be a variable and unknown continuum contamination from nearby sources due to variations in seeing conditions. 

In this paper we measure the correlation between \nad\ EW and extinction in the Milky Way (MW) to a significantly higher precision than achieved before. We show that indeed the correlation is robust based on two different datasets but that the functional shape of the correlation differs significantly from what has been derived before. We discuss the implications of these differences in terms of the inference one can make about the dust content of the MW. We also derive a simple empirical law that allows one to estimate an extinction based on measured \nad\ equivalent widths.

While the correlation between dust extinction and the EW of \nad\ has been shown to exist, it is of value to see how it arises theoretically.
 
By definition, 
\begin{equation}
	{\rm EW} = \int{(1-e^{-\tau(\lambda)})d\lambda},
\end{equation}
which, at low optical depths $\tau$ translates to 
\begin{equation}
	{\rm EW} \propto \tau.
\end{equation}
The optical depth is the column density of the neutral sodium, $N_{\rm Na\,I}$, times the absorption coefficient, which we denote as $\kappa_0$. For a given gas-to-dust ratio $f_{\rm g2d}$, a sodium fraction $f_{\rm Na}$ (the mass fraction of all the gas that is in sodium), and sodium ionization fraction $f_{\rm ion}$, we therefore get
\begin{equation}\label{eq:3}
	{\rm EW} \propto \kappa_0 (1-f_{\rm ion}) f_{\rm Na} f_{\rm g2d} N_{\rm dust}, 
\end{equation}
where $N_{\rm dust}$ is the column density of the dust. 

We next take the definition of the total to selective extinction,
\begin{equation}
	R_{\rm V}=\frac{A_{\rm V}}{E_{\rm B-V}}.  
\end{equation}

$A_{\rm V}$ is proportional to $N_{\rm dust}$. We can assume any arbitrary function, which we will denote as $F$. So
\begin{equation}
N_{\rm dust} = F(R_{\rm V},{E_{\rm B-V}}), 
\end{equation}
which we can finally combine with Equation \ref{eq:3}, to get
\begin{equation}\label{eq:6}
	{\rm EW} \propto \kappa_0 (1-f_{\rm ion}) f_{\rm Na}  f_{\rm g2d} F(R_{\rm V}, {E_{\rm B-V}}). 
\end{equation}

Given the large number of proportionality factors, and how uncertain (and varying) they are in the MW, it is quite astounding that a correlation between EW and \ebv\ survives at all. Complex physical processes in the ISM further muddy the picture as, for example, the sodium fraction will be influenced by depletion  that in turn depends on the dust content. Furthermore, $f_{\rm ion}$ is large, and sensitive to photo-ionization as a result of the low ionization potential of sodium (5.1\,eV). Perhaps what saves the day is a more direct correlation between metal content and dust, bypassing all of the gas-related variables. 

A further complication, which will have significant implications on the shape of the relation, is that a random sight line through a galaxy will often cross more than one gas cloud. This allows the EWs of the lines to grow beyond saturation if one does not possess the resolution to separate the systems. Even if the absorption lines of the different systems can be separated, the extinction incurred by the different clouds cannot be. MZ97 resolve this issue by removing from their sample stars that show multiple absorption systems. This effectively assumes that the instrumental resolution is better than the velocity dispersions of the systems and that those systems that appear single are indeed fully resolved. In this paper we sum the contributions from different components, even when we can separate them, effectively averaging over the typical number of clouds crossed by a sight line.

As we show below, the correlation does exist, with a surprisingly low scatter, yet with a functional structure that reflects somewhat the richness of the processes involved. 

In Section \ref{s:data} we present our samples, which we inter-compare in Section \ref{s:compare}. In Section \ref{s:EWvsSFD} we look at the correlation between \nad\ EW and \ebv, deriving the empirical tool in Section \ref{s:disc}. We conclude in Section \ref{s:conc}.

\section{Data and Measurement}\label{s:data}

Our purpose is to measure the EWs of \nad\ and to compare them to \ebv\ values. For the latter we use the maps of \citet[][hereafter SFD]{schlegel98} who measured the distribution of dust throughout the sky using far infrared emission as a tracer of dust content. Recently \citet{schlafly10} and \citet{schlafly11} have found that the SFD maps may need to be re-normalized as they over-predict extinctions by 14\%. These authors also find that there are offsets between the northern and southern skies as well as higher order corrections in some smaller regions of the sky. However, they do not provide improved maps yet and claim they will do so when more data are available. Since this rescaling is a temporary fix, we do not apply it, and note that it does not affect our result in any significant fashion. A reader who prefers to apply this scaling can simply multiply any color excess value that we derive based on the SFD normalization by $0.86$. 

In order to measure the narrow sodium absorptions we require large numbers of high-resolution observations of background sources. We compile data from two different catalogs, with quite different properties. One sample consists of over a hundred high-resolution spectra of quasars (QSOs), where the analysis of absorption lines is straightforward. A second sample includes hundreds of thousands of low-resolution spectra from SDSS that require more subtle methods. The samples complement each other and allow for various consistency checks. 

\subsection{High-Resolution Sample}
Over the past two decades many tens of QSOs have been observed at high resolution with the Keck telescopes for various science goals. We use here a sample of 117 QSOs that were obtained using the HIRES (43 spectra) and ESI (74 spectra) cameras. This sample includes only spectra that cover the wavelength range of \nad. We henceforth call this combined set the `HIREZ' sample. 

All of the spectra were drawn from the the UCSD/Keck Damped \lya\ Database \citep{prochaska07} and surveys of metal-strong damped \lya\ systems \citep{herbert-fort06} The spectra were reduced with either the MAKEE, HIRedux or ESIRedux packages to optimally extract, calibrate, and coadd the spectra. Each spectrum was continuum normalized and shifted to the heliocentric vacuum reference frame. We refer the reader to those papers for full details on data acquisition and reduction.

\begin{figure}

\includegraphics[width=1\columnwidth]{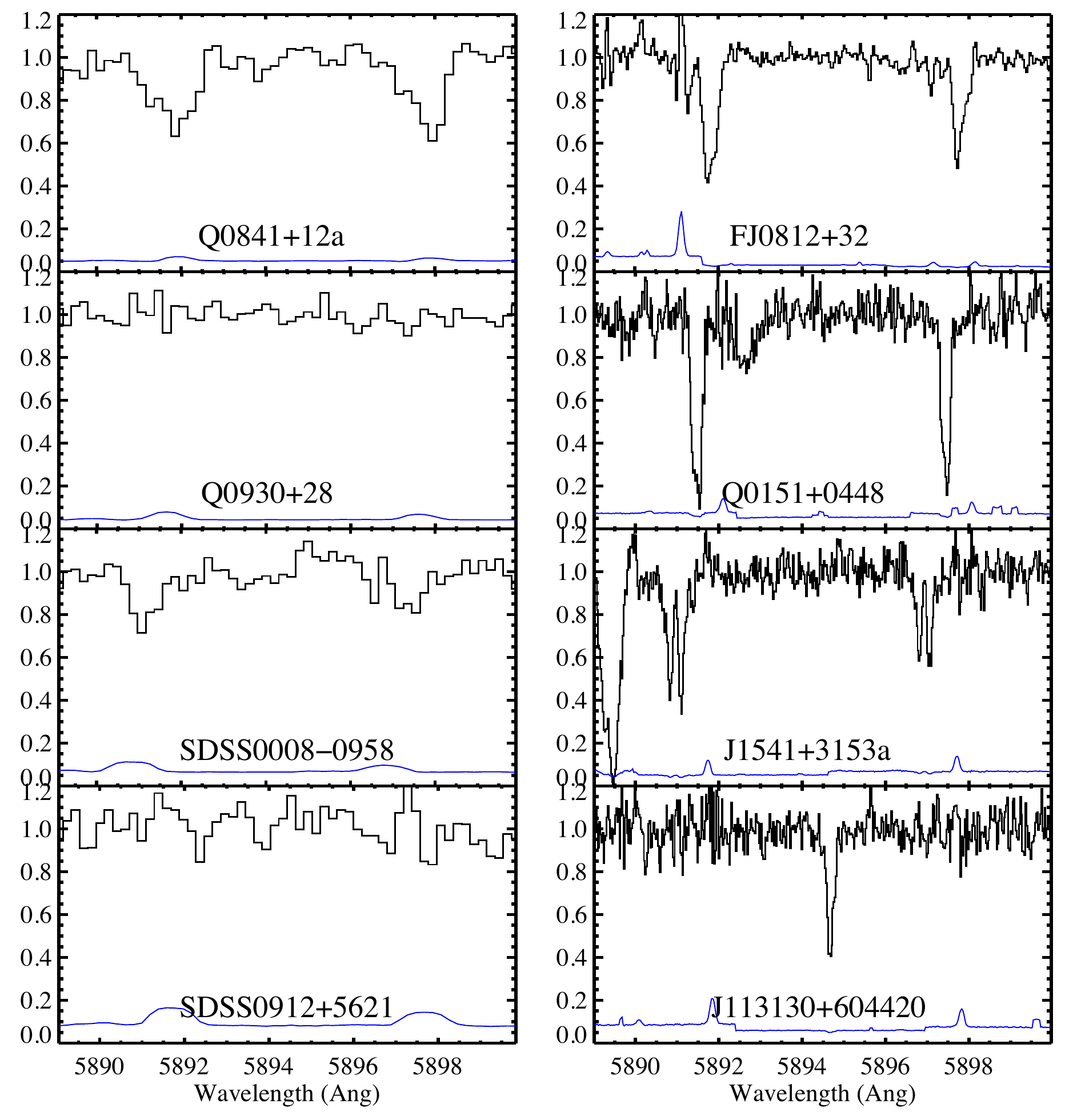}
\caption{Representative samples of Keck/ESI (left) and Keck/HIRES QSO spectra
covering the \nad\ doublet absorption due to the Milky Way.  Each spectrum has
been continuum normalized.  These QSO spectra show occasional
absorption from extragalactic systems in the wavelength region of interest
(e.g.\ at $\lambda \approx 5894.5$\AA) toward J1131+6044.  Such
features were carefully identified and ignored in our analysis. Blue lines are the $1\sigma$ error arrays. 
\label{f:eg_hirez}}
\end{figure}

We examined each spectrum near the wavelength of \nad\ for the presence of the doublet which we could associate with Galactic absorption. Despite the high spectral resolution ($R \approx$ 40,000--50,000 for HIRES, and $R \approx$ 6,000--9,000 for ESI), the data were occasionally compromised by strong \nad\ emission and those few spectra were eliminated from the analysis. Examples of the spectra that were considered is presented in Figure~\ref{f:eg_hirez}.

For the spectra that were free of excessive sky contamination, we measured the equivalent width of each member of the doublet using boxcar summation and calculated the corresponding error, summing over multiple systems whenever they are resolved. Table~\ref{t:joint} summarizes these measurements. 
We caution that the error estimates for these measurements ignore the systematic error of continuum placement which imposes an additional $\approx 5-10\%$ uncertainty. Due to the poorer resolution of the instrument, the ESI measurements typically have larger scatter and uncertainty.

\subsection{SDSS Sample}

SDSS observed more than $100,000$ QSOs, and nearly a million galaxies, albeit with low resolution ($R \approx 1800$) and often with quite low signal to noise ratio (S/N). Our sample includes all the spectra in the 8th data release of SDSS, though we use only spectra with redshifts $z>0.005$ to mitigate any contamination from \nad\ absorption in the rest-frames of the QSOs and galaxies. After cleaning out a few spectra of very low quality, we remain with $121,391$ QSOs and $860,883$ galaxies, nearly one million spectra in total, to which we refer as the SDSS sample. 

In order to detect and measure the \nad\ line in the SDSS spectra they must be combined in large numbers (typically more than 200; we test this throughout this paper). We attempted to measure the EWs on individual SDSS spectra, so as not to resort to binning. However for the vast majority of spectra we could not get a double gaussian fit to converge and give a useful measurement. We therefore create high-S/N spectra by means of various stacking schemes. First we linearly interpolate every spectrum to an identical and denser wavelength grid in the range $5800-6000$\,\AA. We then divide the spectra by a low order polynomial in order to remove continua of any shape and to normalize flux levels. We group them with different binning rules and last, combine them using median filtering. Since the \nad\ lines are often weak and always blended, we measure the EW by fitting a double Gaussian to the relevant wavelength region. This is a two step process where we first fit for the stronger $D_2$ line (restricting our fit to wavelengths smaller than 5894\,\AA; fitting for the line center, width, and amplitude), we then subtract the best fit Gaussian from the spectrum, and finally similarly fit for the $D_1$ line (at $\lambda>5984$\,\AA). We find that this method robustly recovers the doublet at all detectable strengths. 

\begin{figure}
	\center
\includegraphics[width=1\columnwidth]{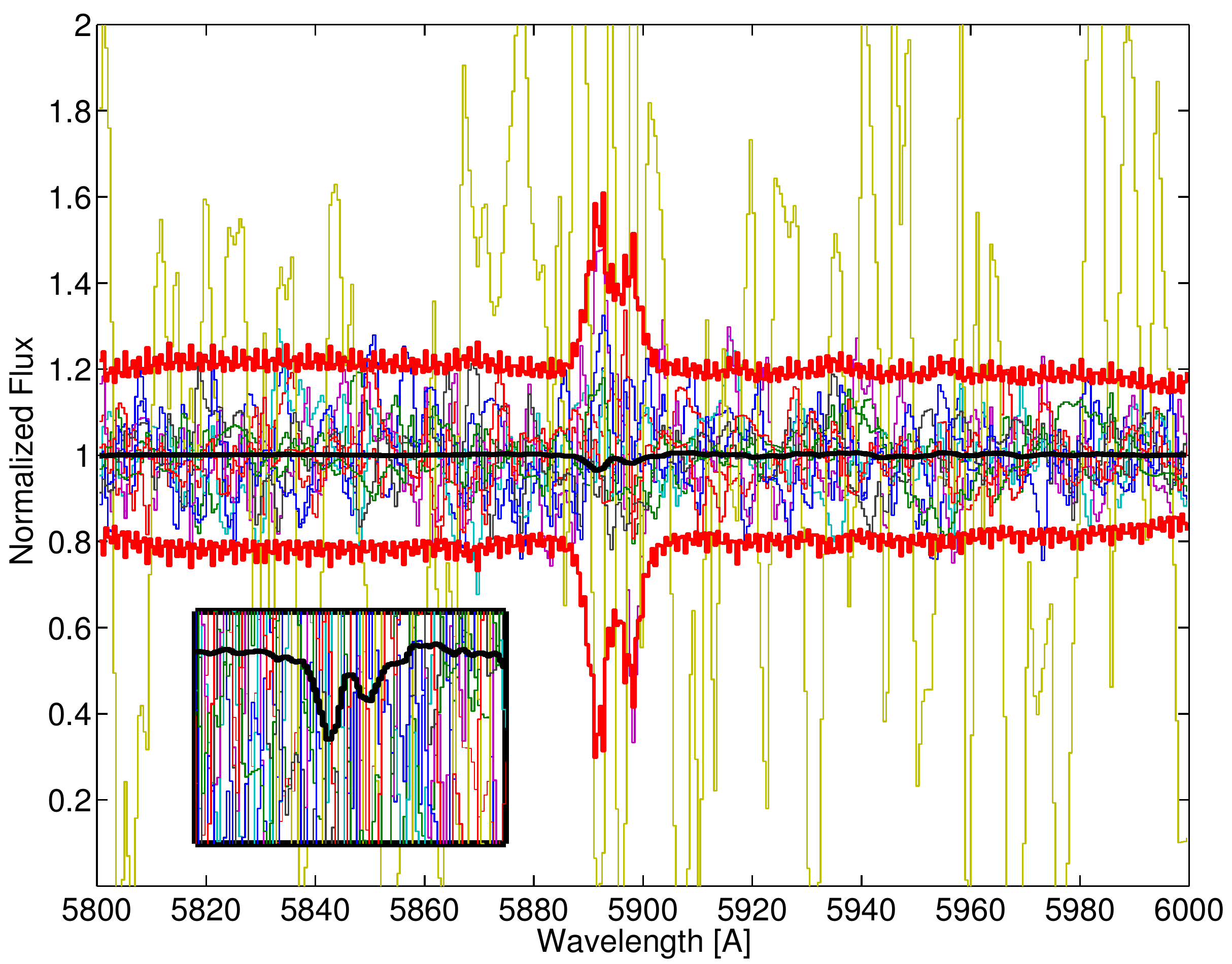}
\caption{We median combine \textbf{all} of the (normalized) SDSS spectra (nearly a million), the resulting spectrum with a clear detection of the \nad\ doublet is plotted with a thick black line. A random subset of ten input spectra are shown in the background in various colors, in order to demonstrate their noise properties. In thick red lines we show the bottom 5 and top 95 percentile curves.\label{f:combo}}
\end{figure}

In Figure \ref{f:combo} one can see the effect of combining all the SDSS spectra, though only 10 random ones are shown in the background for clarity, as well as the 5 and 95 percentile curves. Despite the large amplitude noise in the spectra (in fact at our wavelength of interest the noise is even greater due to residual contamination from atmospheric \nad\ lines) we recover a strong doublet at the \nad\ wavelengths. The resulting spectrum has the theoretical resolution of the SDSS spectrograph, of order $2.5$\,\AA\ at this wavelength. 

\section{ HIREZ vs. SDSS: comparing the samples}\label{s:compare}

\begin{figure}
	\center
\includegraphics[width=1\columnwidth]{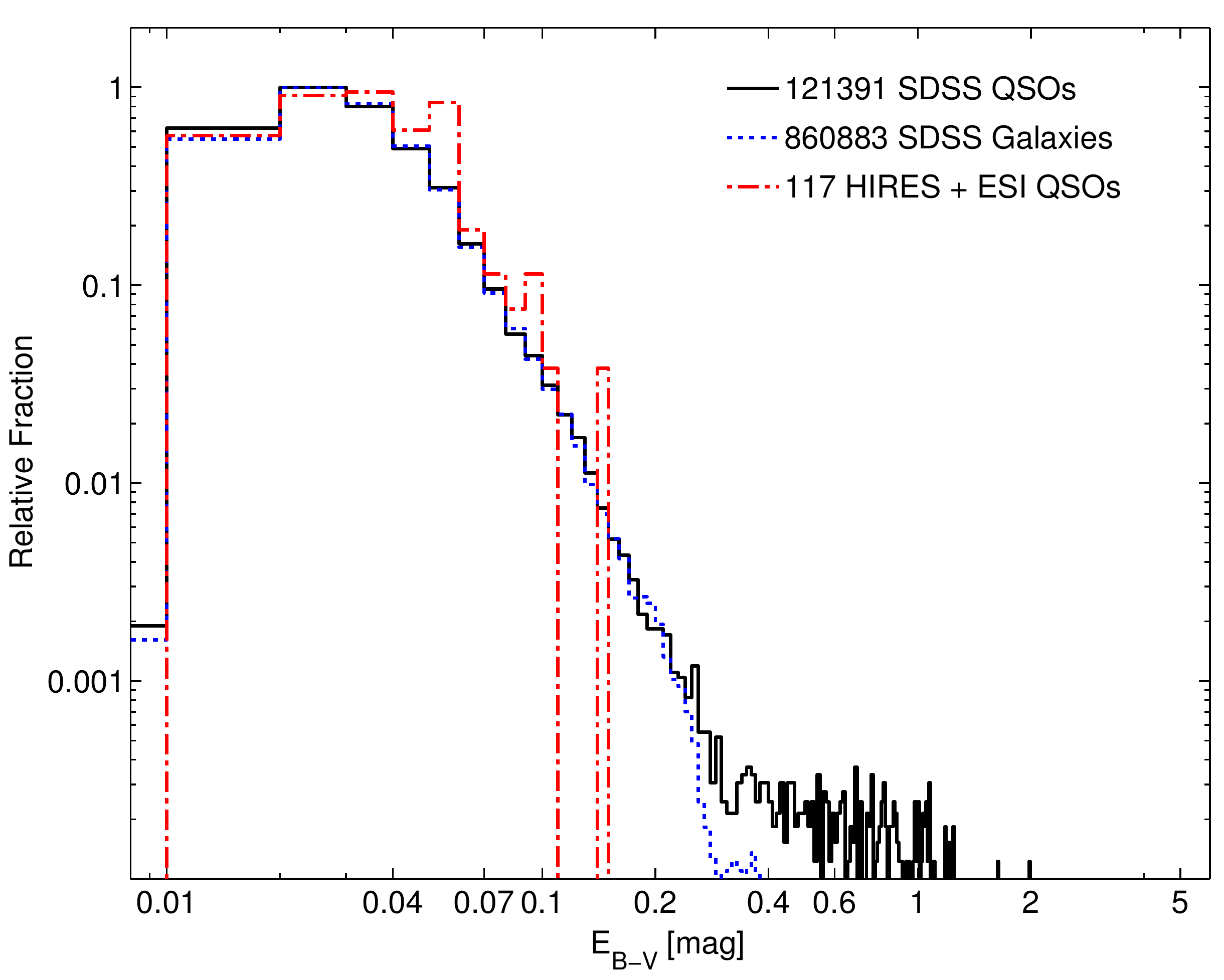}
\caption{\ebv\ distribution for our samples of spectra, SDSS QSOs in black, SDSS galaxies in blue, and HIREZ in red. Clearly, these samples are similar, and mostly probe the low extinction range, however the large number of SDSS spectra does include more highly reddened objects. \label{f:histEBmV}}
\end{figure}

By construction, both samples are typically at high Galactic latitude and low Galactic extinctions, as derived from the SFD maps. Figure \ref{f:histEBmV} shows the respective distributions in \ebv. The distributions match well (which we confirm by a Kolmogorov-Smirnov test), and mostly probe the region of \ebv$<0.1$. However within the gargantuan number of SDSS spectra there is a non negligible fraction at higher extinctions, up to \ebv$\sim 6$. 

In order to assess whether the \nad\ EWs for both samples are consistent and to determine the scatter in the measurements, we compare the samples directly as follows. For every set of coordinates of a HIREZ object we would like to stack SDSS spectra of targets in a similar position in the sky. The maximum angular separation is a compromise between two effects. A larger separation would improve the S/N in the combined spectrum, at the cost of larger scatter due to variance introduced by averaging over too different sight lines. Accordingly, a smaller separation will reduce significantly the number of sources, and limit our ability to measure the EW.

\begin{figure}
	\center
\includegraphics[width=1\columnwidth]{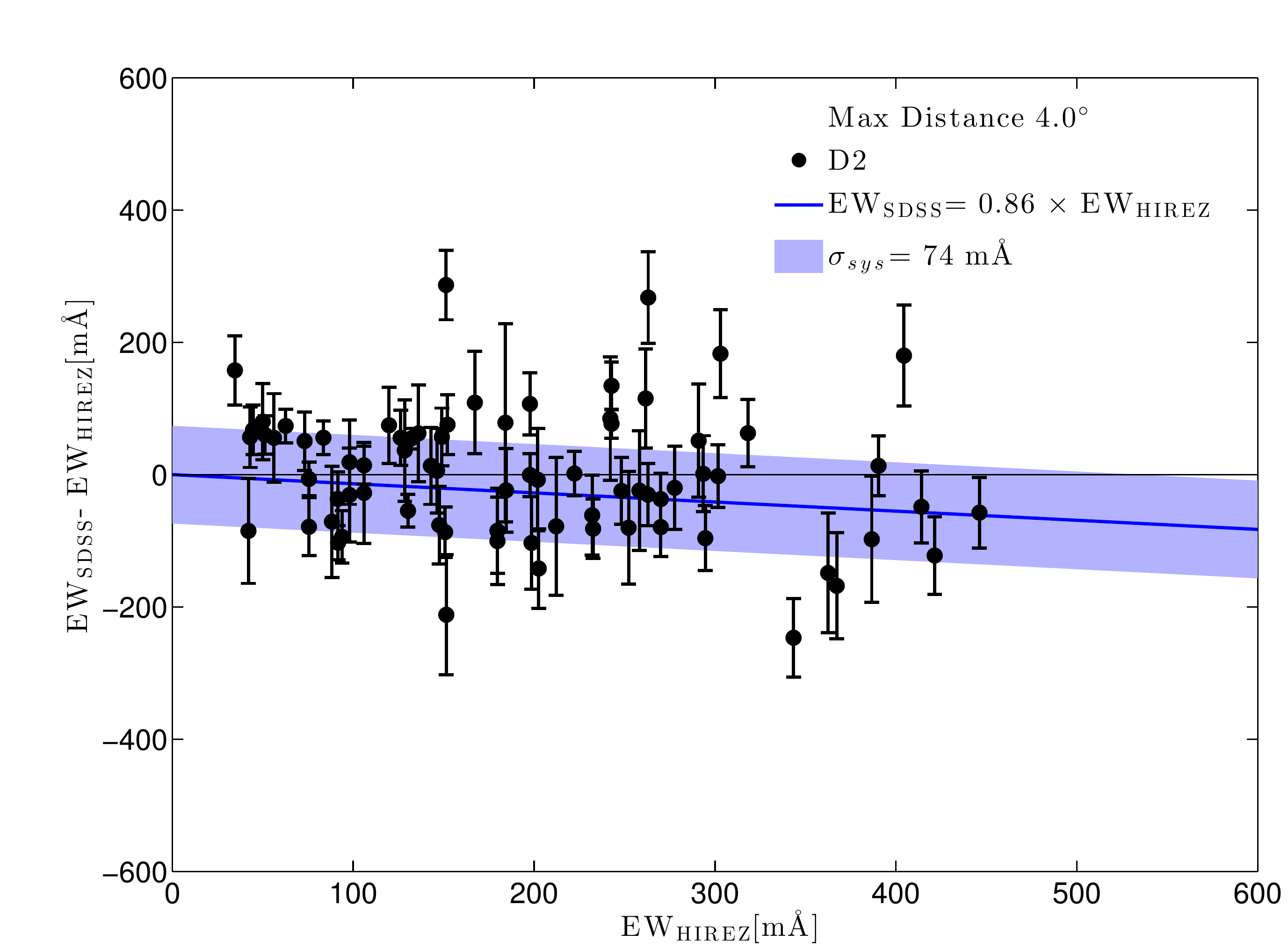}
\caption{We compare the EWs derived for both samples at a similar position on the sky. Plotted here are the residuals between the measured EWs of the stronger $D_2$ line for a maximal separation of $4^{\circ}$ between binned SDSS spectra. We find a systematic uncertainty $\sigma_{sys}$ of order 70\,m\AA, and that stacking introduces a bias of about 10\% as discussed in the text. Residuals in black circles, best fit line in blue, with $\sigma_{sys}$ shaded.\label{f:bias}}
\end{figure}

Requiring at least 200 SDSS objects per bin (otherwise the combined spectra are too noisy), and trying a range of thresholds, we measure the EW of the (stronger) $D_2$ line and compare the EWs of the HIREZ spectra versus stacked SDSS EWs, and perform a linear fit. The slope between the two samples (which should be of order unity) is near 0.9 for maximal separations (i.e., the diameters of the circular regions around each HIREZ object) of $2-4^{\circ}$. Greater separations show a larger offset, as expected, up to about 20\%. This scale is consistent with our measurement of the typical separation on the sky over which the SFD maps change significantly: for our fields (effectively the SDSS footprint) a standard deviation of 0.01\,mag in \ebv\ occurs within a distance of $2^{\circ}$ (see however some works that measure variations on much smaller scales: \citealt{frail94,meyer99,andrews01}). Quite independently of distance, the systematic uncertainty (i.e., the added uncertainty required for $\chi^2$ per degree of freedom (d.o.f) $\sim 1$) is about 70\,m\AA. Figure \ref{f:bias} shows an example for a separation threshold of $4^{\circ}$, though plotted in term of the residuals between the samples, for clarity.

The procedure above supplies us with two numbers. First, our systematic uncertainty is of order 70\,m\AA. A few sources contribute to this number: variations on small scales of the \nad\ absorption, the effects of binning, sky subtraction errors, and perhaps underestimated measurement uncertainties. Second, we find a systematic shift, where measurements on the SDSS spectra are consistently underestimated by about 10\%. This could be expected because of the following effect. As we bin and combine multiple SDSS spectra with a skewed distribution of extinctions, the unavoidable pollution of one bin by members from adjacent bins is asymmetric. The dominant population contaminates the remainder more significantly than vis versa. Since most of our spectra suffer from low-extinction and small EW, any contamination will tend to dampen the EW measured. 

This explains why at large threshold distances the bias grows, as more and more different (and uncorrelated) sight lines are combined. This also entails a bias when we bin spectra according to their \ebv\ values, as derived from the SFD maps. Since these maps have a typical scatter of 0.01\,mag we expect a bias of 10\%. We examine this bias further in Section \ref{s:bias}. In order to correct for this bias, we add a 10\% correction to every EW value derived from SDSS stacked spectra. This correction is small enough to have little effect on our qualitative and quantitive findings.

\section{ EW vs. \ebv\ from SFD Map}\label{s:EWvsSFD}
We next turn our attention to comparing the \nad\ EWs as derived from both datasets to the extinction posited by the SFD maps.

\subsection{HIREZ sample}

\begin{figure}
	\center
\includegraphics[width=.99\columnwidth]{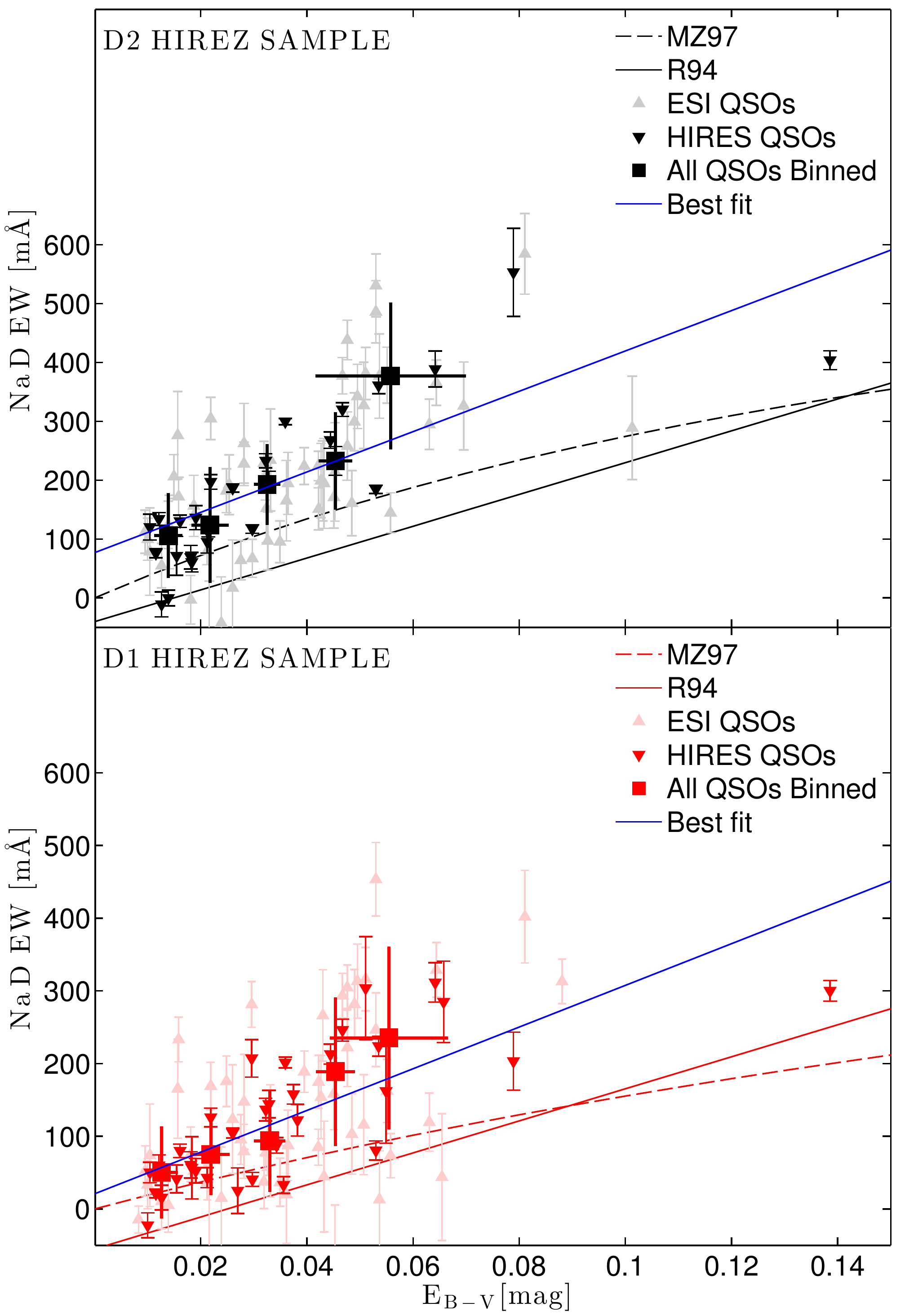}
\caption{\nad\ EW vs. \ebv\ for the HIREZ sample (bottom: the D$_1$ line, top: D$_2$). The correlation is most apparent for the binned data or for the HIRES data, since the ESI EWs are much noisier. However the correlation is significant and consistent across the subsamples. Our best fit linear curve (blue line) clearly diverges from the derivations of MZ97 (dashed lines) and R94 (solid lines).\label{f:ew_ebv_hirez}}
\end{figure}

Figure \ref{f:ew_ebv_hirez} shows the EWs as a function of \ebv\ for the HIREZ sample. There is a clear correlation between those two values (Pearson's correlation coefficients of 0.7 and 0.6 for the $D_2$ and $D_1$ lines respectively, with $p$ values smaller than $10^{-10}$). The correlation is more readily apparent to the eye when one bins the data over \ebv, where the bin edges are set to have about 20 measurements per bin. While robust, the correlation does not follow the trend we expect from the curves of either R94 or MZ97\footnote{Note that MZ97 do not actually fit for the \nadone\ correlation. They discuss the fact that \nadtwo\ should be twice as strong due to ratio of oscillator strengths but that the lines should get closer in strength as the stronger line saturates. They find that their data corroborates this, asymptotically converging toward a ratio of 1.1. We therefore simulate the MZ97 \nadone\ curve by modulating the \nadtwo\ best fit of MZ97 with a function that starts at 2 and exponentially reaches a value of 1.1.}. If we assume a linear relation between \nad\ EW and \ebv\ in that range of extinctions, our best fit slope is about twice larger than the slope of the MZ97 curve. The systematic scatter is of order 50\,m\AA.

\subsection{SDSS sample}

We next turn our attention to the SDSS sample, where we bin the spectra over similar \ebv\ values as derived from the SFD maps. Here again the number and sizes of bins were not determined arbitrarily. We tested various procedures that all gave consistent results. For  $0.02 < $\ebv$<0.3$, where we have an abundance of spectra and signal, rather than fixing the sizes of the bins, we fix the number of objects per bin. The number of bins for this extinction range was determined by varying it from 5 to 200, every time fitting a quadratic function to the resulting EW vs. \ebv. We found that the results are always consistent and stable and do not depend on the number of bins. However, outside this range of extinctions, we only have about 1500 spectra at \ebv$>0.3$, spanning a large range of extinctions -- going to \ebv$\sim 6$ (with only 54 in the range $2<$\ebv$<6$). For these we therefore set additional bin edges at color excesses of 0.5, 1, and 2. All the spectra at \ebv$<0.02$, where the EW is very small and difficult to measure, are stacked in a single bin. We present below results using a total of 50 bins that seem to well sample the relation.

\begin{figure}
	\center
\includegraphics[width=.98\columnwidth]{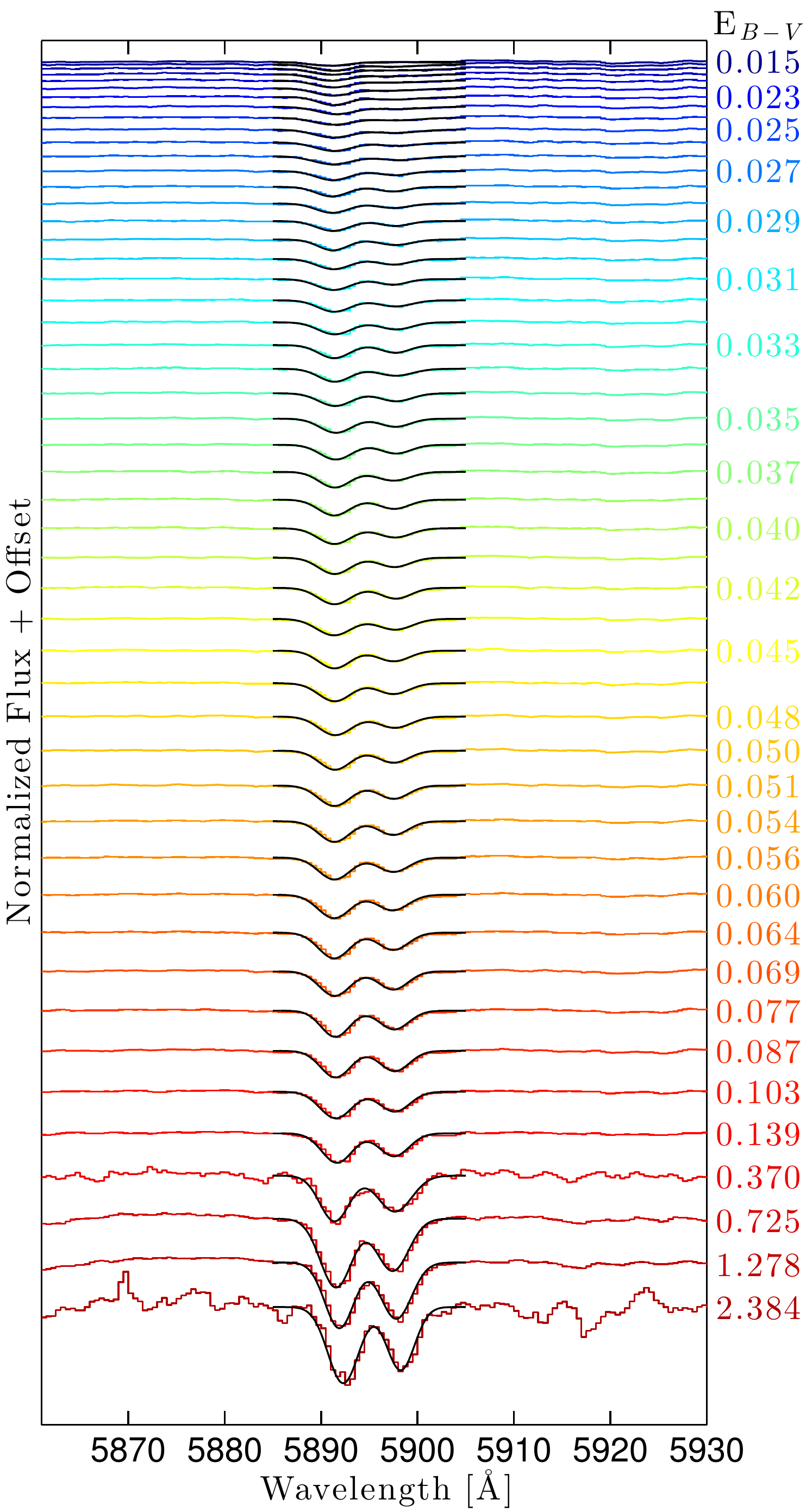}
\caption{Binned SDSS spectra sorted from low to high extinction (top to bottom), superposed with the best fitting double Gaussian (black). The strong correlation between the EW of \nad\ and extinction is clearly apparent to the unaided eye.\label{f:res_ew}}
\end{figure}

Figure \ref{f:res_ew} shows the 50 binned SDSS spectra, as well as the fit to the \nad\ lines, sorted from low extinction (top) to high extinction (bottom). Clearly the spectra have extremely high S/N, deteriorating as we reach the highest bins where objects are scarcer. The doublet is well detected, and apparently the double Gaussian fit works well, and recovers the line shapes. Another obvious result is the clear correlation between the EW of both lines and extinction. One can follow the curve of growth by eye, as $D_2$ gets stronger first, stalls and is nearly caught up with by the weaker $D_1$ line that saturates later.

\begin{figure}
	\center
\includegraphics[width=.99\columnwidth]{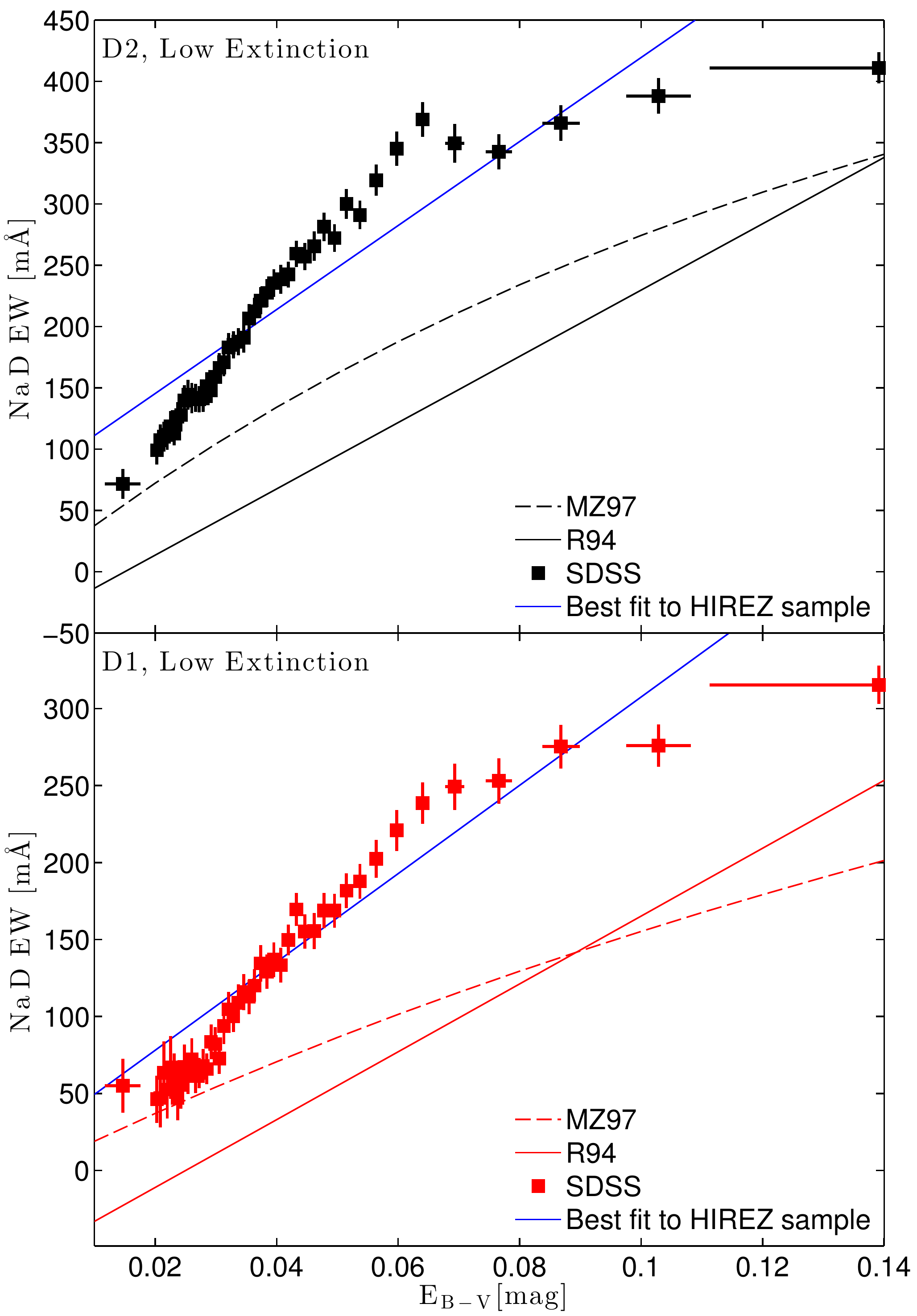}
\caption{\nad\ EW vs. \ebv\ for the SDSS sample (squares; bottom: D$_1$, top: D$_2$, at low extinctions. While there is a clear correlation consistent with the one derived from the HIREZ sample (blue line), it again diverges from the curves derived by MZ97 (dashed lines) and R94 (solid lines), with an intricate nonlinear  structure.\label{f:ew_ebv_sdss_I}}
\end{figure}

\begin{figure}
	\center
\includegraphics[width=.99\columnwidth]{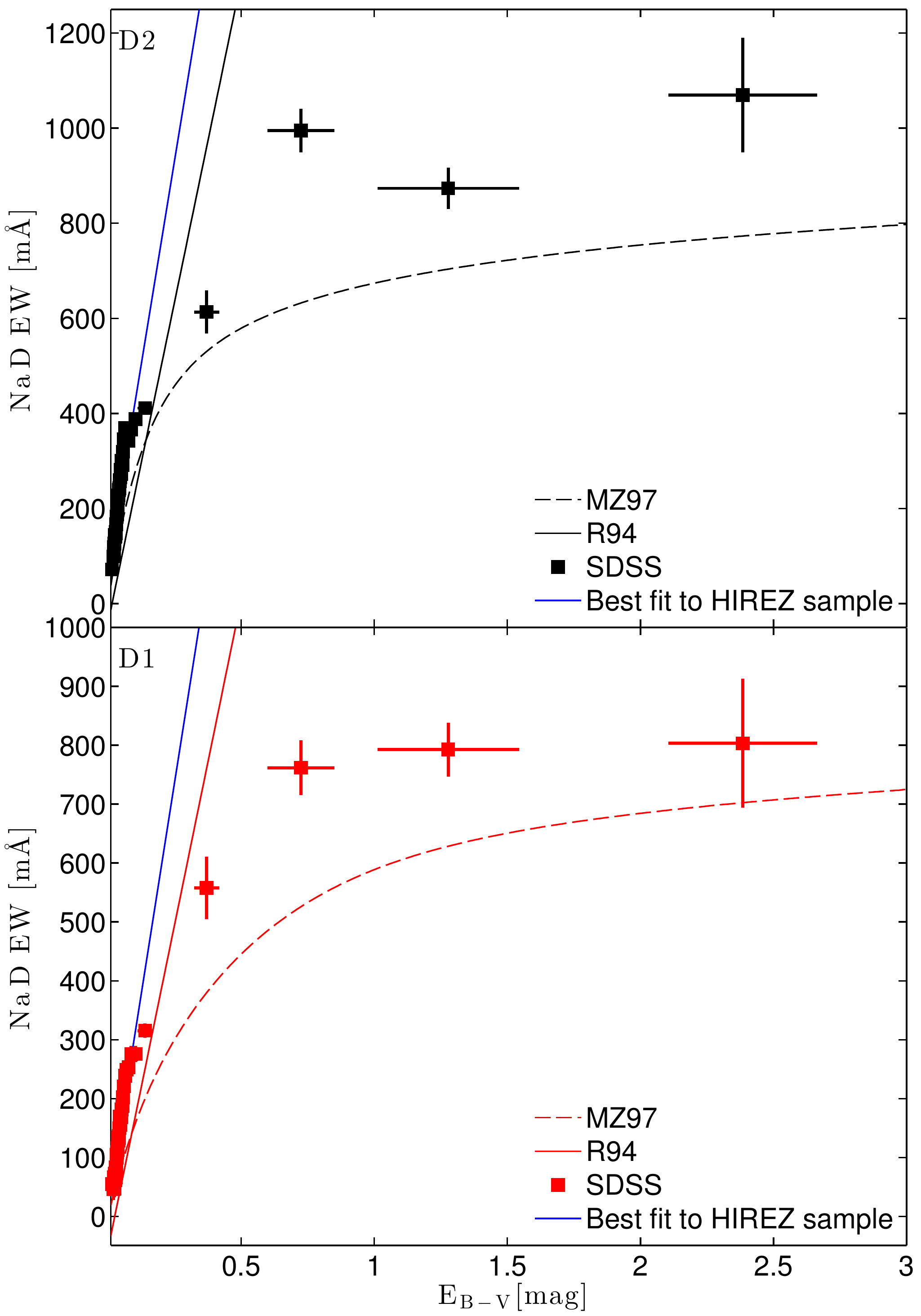}
\caption{Same as Figure \ref{f:ew_ebv_sdss_I}, for the full range of extinctions.\label{f:ew_ebv_sdss_II}}
\end{figure}

In Figures \ref{f:ew_ebv_sdss_I} and \ref{f:ew_ebv_sdss_II} we show the EW vs \ebv\ at low extinctions, and over the full range that we probe, respectively. There are quite a few remarkable features in these plots. First, the correlation is very tight, with a similar slope at low extinction to the one derived from the HIREZ sample. This is true for both \nad\ lines. Second, similarly to what we find for the HIREZ sample, but even more so here, there is a very clear discrepancy between the curves derived by R94 or MZ97 and our findings. Last, while the correlation is quite linear up to extinctions of about 0.06, it then flattens, reaching a first plateau around 350\,m\AA\ (D$_2$; 270\,m\AA\ for D$_1$), but with a much more striking saturation feature occurring at \ebv$>0.7$, and an EW near 1000\,m\AA\ (D$_2$; 800\,m\AA\ for D$_1$), as readily apparent in Figure \ref{f:ew_ebv_sdss_II}.

If we combine all the SDSS spectra in one bin the resulting spectrum has EWs of 160\,m\AA\ and 90\,m\AA\ for the \nad\ lines. At a median \ebv\ of 0.026 this is consistent with the results of the `less aggressively binned' data.

\section{Discussion}\label{s:disc}
\subsection{Comparison to previous results}

Since the data and analysis of MZ97 are more relevant at the extinction ranges we probe, in addition to their curves being less discrepant from our data, we hereafter focus on these rather than on the results of R94. 

Our data seem to point to a rather simple multiplicative re-scaling of the MZ97 curves. We perform joint fits of our two (binned) samples to the best fitting MZ97 curves for both \nad\ lines and find that models where EW$\rightarrow n_1$EW 
or \ebv$\rightarrow n_2$\ebv\ bring the MZ97 curves to our data (or vis versa). A systematic uncertainty of order 30\,m\AA\ suffices to reach $\chi^2/dof = 1$.

The best fit requires the EWs to be multiplied by about 1.5. Alternatively, to \ebv\ we need to apply a factor of 2. This result is stable, whether we fit only one sample or the other, one absorption feature or the other. 
A value of 1 for the stretch (i.e., no change) is  $>5\sigma$ away from our best fit. 

We find it unlikely that MZ97 have underestimated their EW measurements by a factor of $1.5$, considering the fact that they span such large extinctions (up to \ebv$=1.5$), with commensurately large EWs (up to 600\,m\AA). In view of the strength of the signal, in two independent datasets with different methods, we find it equally improbable that we overestimated ours by that same amount. 

Alternatively MZ97 may have overestimated their \ebv\ values by a factor of 2, or  that the SFD extinctions we use are as severely underestimated. Here again neither explanation is very attractive. 

The only remaining possibility is that somehow we are not comparing `apples to apples', by probing a different ISM from the one MZ97 probe on average. Restricting ourselves initially to low optical depths, we find a much steeper slope than the one observed by MZ97. This could be explained if we allow a systematic change between the samples in one of the parameters of our Equation \ref{eq:6}, namely, $\kappa_0$, $f_{\rm ion}$, $f_{\rm Na}$, $f_{\rm g2d}$, or $R_{\rm V}$. Obviously all these factors are not well measured, known, or understood, except for some very specific environments. However, we can explore qualitative arguments on the amount and direction of change we can expect when moving from the low galactic latitude, disk environment of the MZ97 stars to the high galactic latitude, bulge-like gas and dust we that we are mostly observing. 

Perhaps the simplest parameter is $R_{\rm V}$. In the denser gas- and dust-rich star forming environments of the MZ97 stars we expect larger grains to be more easily built. Larger grains lead to larger values of $R_{\rm V}$ and therefore this effect would work in a direction opposed to what we measure. 

An additional dependency is similarly ruled out from simple arguments. The Galactic metallically gradient would reduce $f_{\rm Na}$ in our sample when compared to more central regions. This effect would therefore work in the wrong direction. 

The three remaining parameters are the ionization fraction, variations in the gas-to-dust ratios, or changes in the depletion of sodium. The ionization fraction will depends on the radiation field, which in turn can vary significantly in different regions. In the denser disk of the galaxy we expect larger UV fluxes, and a greater ionization fraction.  According to \citet{savage96}, one finds a larger gas-to-dust ratio in the halo. Last, in dust rich environments we perhaps expect greater depletion of the sodium from the gas to the solid phase. All of these effects will make $f_{Na}$ greater for our sample, so could explain what we measure. However, they would more markedly affect high extinction regions, which is not what we measure -- an overall stretch of the correlation. There is also evidence that sodium depletion is rather constant and independent of environment \citep{phillips84}. 

We checked whether the correlation depends on the Galactic latitude by breaking the SDSS sample to two subsamples, at absolute Galactic latitudes above or below $50^\circ$. We find no measurable difference between the two correlations. While one does expect differences in the dust distribution and properties as a function of galactic latitude, it seems that binning the data in \ebv\ erases the distinction between the environments. 

The answer probably lies in the different treatment of multiple line systems. MZ97 make a point of using solely those stars towards which they resolve only single systems, and they reject from their fit stars that show more than one set of absorption lines. As expected, the stars they reject tend to have larger EW since more clouds allow for more absorption to happen at a different velocity, circumventing saturation. However, our sample lies outside the Galaxy, and our lines of sight cross multiple clouds on average. This is also the most likely scenario when one observes an extra-galactic source and wants to correct for host (or MW) extinction. This, at least qualitatively, explains why we tend to measure a steeper correlation. Indeed our data and best fit relation seem consistent with the handful of multiple-line systems that MZ97 reject from their analysis. 

Whether individual systems are resolved or not depends on the instrumental resolution as well as the velocity dispersion in the clouds. Many of our spectra observed with the HIRES spectrograph clearly show multiple systems, often separated in velocity by a few km\,s$^{-1}$, while most of the rest have complex line shapes that are neither Gaussian nor Lorentzian. This can be seen in Figure \ref{f:eg_hirez}. We assume the same applies for the ESI and SDSS spectra that do not offer the resolution or the S/N to assess this. For that reason and for consistency between the samples, we add up the EWs whenever multiple systems are resolved in the HIREZ sample. 

This also sheds light on the functional form of the correlation we find. The first `plateau' in the correlation can be attributed to saturation from single clouds. For a Doppler parameter $b \approx 5 \, \rm km \, s^{-1}$, the \nad\ doublet should indeed begin to saturate at $\approx 200$\,m\AA.

However,  most of our sight lines across the MW will traverse multiple clouds. These individual systems will not be resolved in our SDSS spectra, and can contribute to the measured EW, allowing the curve of growth to continue rising past saturation. The second, stronger saturation, that occurs at EWs near 1000\,m\AA\ (D$_2$; 800\,m\AA\ for D$_1$) might therefore be the result of all these clouds saturating. Since our method averages over multiple sight lines, the ratio between the two saturation levels is the average number of such clouds along a typical line of sight, about three. 

\subsection{Empirical Relation and Practical Guide}

\begin{figure}
	\center
\includegraphics[width=.99\columnwidth]{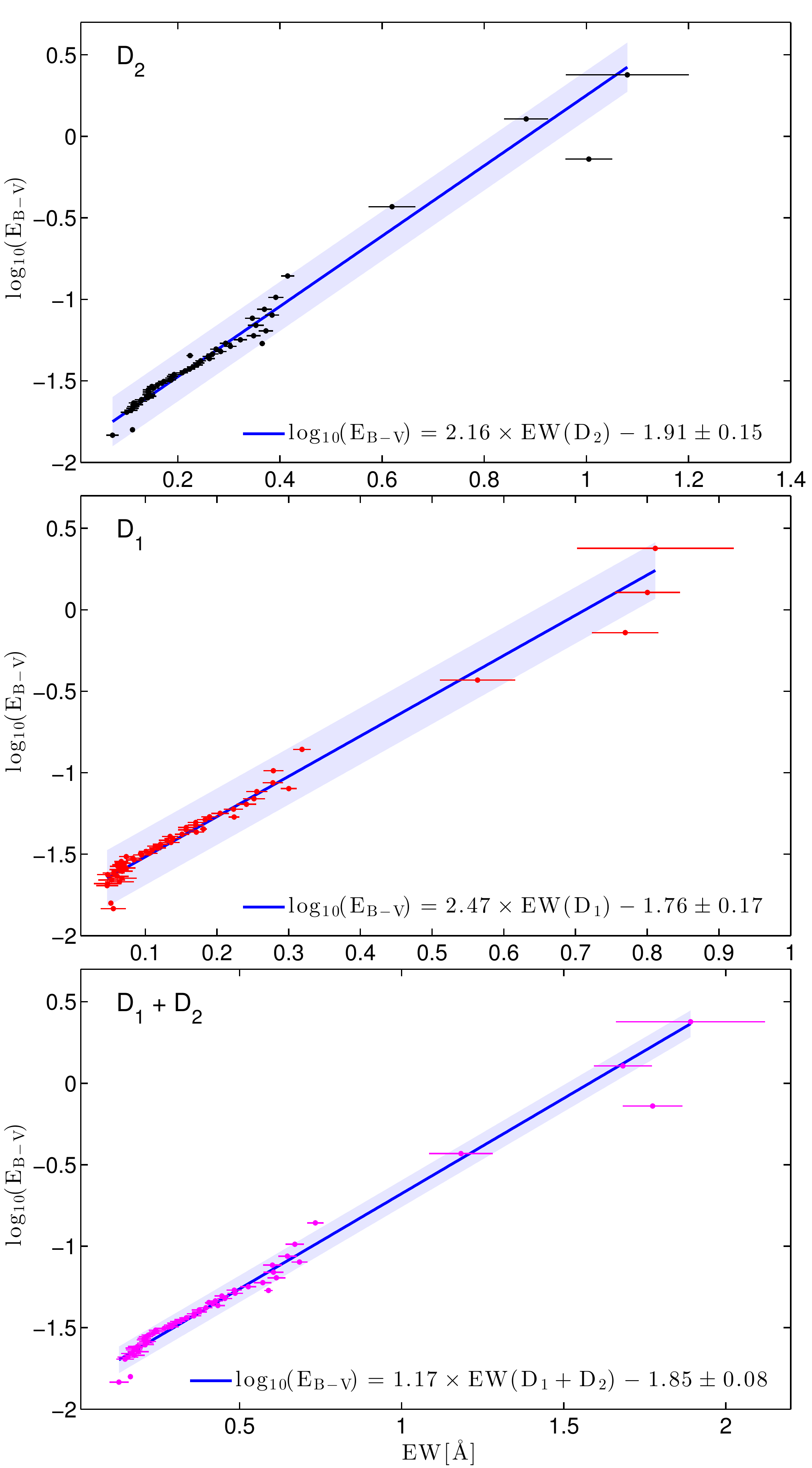}
\caption{$\log_{10}$(\ebv) vs. \nad\ EW for both samples merged. 
D$_2$ top, D$_1$ middle, and their sum in the bottom panel. The correlation can be well approximated with a linear fit when one takes the logarithm of the color excess. The light blue band marks the 1$\sigma$ uncertainty which is dominated by the intrinsic scatter in the correlation. \label{f:empirical}}
\end{figure}

We aim to provide the reader with the simplest yet precise description of the correlation, in order to derive extinctions from measured \nad\ lines. Having tested various functions to fit the data, we found that the logarithm of the color excess is, within the uncertainties, quite linear with respect to the equivalent widths. This is shown in Figure \ref{f:empirical}, where the relations to both lines (and their sum, in case they are not resolved) are shown. Note that for convenience we switch the axes when compared to previous figures, and use \AA\ units for the EW rather than m\AA. The uncertainties in the derived parameters are dominated by the systematic uncertainty of 70\,m\AA\ which we propagate. This should encompass the natural scatter in the correlation and entails an uncertainty in derived color excesses of 30--40\%. The relation for the $D_2$ line is:
\begin{equation}
	\begin{split}
\log_{10}(E_{\rm B-V}) =& 2.16\times{\rm EW(D_2)} - 1.91 \pm 0.15, 
\end{split}
\end{equation}
for the $D_1$ line: 
\begin{equation}
\begin{split}
\log_{10}(E_{\rm B-V}) =& 2.47\times{\rm EW(D_1)} - 1.76 \pm 0.17, 
\end{split}
\end{equation}
and for a measurement of both lines combined:
\begin{equation}
\begin{split}
	\log_{10}(E_{\rm B-V}) =& 1.17\times{\rm EW(D_1+D_2)} - 1.85 \pm 0.08.
\end{split}
\end{equation}

There are a noteworthy caveats regarding these relations. First, for Galactic (or very low redshift) sources the contamination from atmospheric sodium might be difficult to remove especially in lower-resolution spectra. 

Second, if there are indications that the extinction of a given source is  dominated by circumstellar dust rather than a diffuse interstellar component there is no guarantee that the extinction will follow the relation we derived. Such dust may have a different composition, gas-to-dust ratio, etc. Similarly, our derivation is based on the MW, and therefore implicitly assumes an average $R_V=3.1$. This may not apply in Galactic regions or external galaxies where one expects a different extinction law, with different dust properties.

%
%
%

\section{Conclusions}\label{s:conc}

One would expect wide variations between environments in the parameters that link the column density of neutral sodium and the absorption and scattering properties of dust. Yet, we have found a strong correlation between the EW of \nad\ and dust extinction with a typical scatter below 100\,m\AA.

The relation has a functional shape that differs from previous determinations. We find that this is likely due to the fact that previous studies have only considered stars behind single dust clouds, while extragalactic sight lines will usually cross a number of clouds. Our measurement is the most precise to date, and should serve as a useful tool to correct for the effects of dust on the observations of extragalactic sources, whenever high-resolution spectroscopy exists, or can be obtained. This has already been accomplished for the case of SN2012aw (Van Dyk et al. in preparation) where extinction estimates show that the supernova suffers from little extinction, while the progenitor star was surrounded by a significant amount of dust, showing that the supernova has destroyed the dust surrounding it.

We further have studied the shape of the relation, which sheds light on the structure of the MW. We show for example that within the SDSS footprint, i.e., mostly outside the disk of the Galaxy, a typical sight line crosses about three dust clouds within the MW. 

\section*{Acknowledgments}

In addition to E. Schlafly, our thoughtful and swift referee, we thank the following (alphabetical, probably non-exhaustive) list of people for useful discussions, ideas, and comments on the manuscript:
A. V. Filippenko, 
O. Gnat, 
A. Loeb, 
D. Maoz, 
A.A. Miller,
P.E. Nugent, 
E.O. Ofek, 
J.A. Rice, 
B.P. Schmidt,
M. Sullivan,
B. Trakhtenbrot. 

Some of the data presented herein were obtained at the W.M. Keck Observatory, which is operated as a scientific partnership among the California Institute of Technology, the University of California and the National Aeronautics and Space Administration. The Observatory was made possible by the generous financial support of the W.M. Keck Foundation. The authors wish to recognize and acknowledge the very significant cultural role and reverence that the summit of Mauna Kea has always had within the indigenous Hawaiian community.  We are most fortunate to have the opportunity to conduct observations from this mountain.

Some the data presented were obtained from the SDSS III archive. Funding for SDSS-III has been provided by the Alfred P. Sloan Foundation, the Participating Institutions, the National Science Foundation, and the U.S. Department of Energy Office of Science. The SDSS-III web site is http://www.sdss3.org/. SDSS-III is managed by the Astrophysical Research Consortium for the Participating Institutions of the SDSS-III Collaboration including the University of Arizona, the Brazilian Participation Group, Brookhaven National Laboratory, University of Cambridge, Carnegie Mellon University, University of Florida, the French Participation Group, the German Participation Group, Harvard University, the Instituto de Astrofisica de Canarias, the Michigan State/Notre Dame/JINA Participation Group, Johns Hopkins University, Lawrence Berkeley National Laboratory, Max Planck Institute for Astrophysics, New Mexico State University, New York University, Ohio State University, Pennsylvania State University, University of Portsmouth, Princeton University, the Spanish Participation Group, University of Tokyo, University of Utah, Vanderbilt University, University of Virginia, University of Washington, and Yale University.

\bibliographystyle{mn2e} 
\bibliography{myBIBTeX}
\clearpage
\begin{deluxetable}{lcccc}
\tablewidth{0pc}
\tablecaption{ESI and HIRES MEASUREMENTS\label{t:joint}}
\tablehead{
\colhead{QSO} 
& \colhead{RA}
& \colhead{DEC} 
& \colhead{\nadtwo\ EW}
& \colhead{\nadone\ EW}\\
& (deg) & (deg) & (m\AA) & (m\AA) }
\startdata
\textbf{ESI}\\
\tableline
CTQ460       & 159.7892& -23.2239&$131.0\pm 39.4$&$ 20.3\pm 35.8$\\
FJ0812+32    & 123.0000&  32.0000&$223.9\pm 31.4$&$188.8\pm 28.6$\\
FJ2334-09    & 353.6935&  -9.1366&$303.6\pm 33.2$&$281.3\pm 31.4$\\
HS1132+22    & 173.7835&  22.4519&$ 79.5\pm 23.2$&$ 36.2\pm 23.7$\\
J0121+0027   &  20.4489&   0.4552&$234.2\pm 86.6$&$ 93.6\pm 77.3$\\
J0825+5127a  & 126.3966&  51.4518&$438.1\pm 33.8$&$304.6\pm 31.0$\\
J1017+6116a  & 154.3578&  61.2743&$...$&$-14.7\pm 18.6$\\
J1200+4015a  & 180.1659&  40.2656&$304.8\pm 35.9$&$169.0\pm 32.7$\\
J1238+3437a  & 189.6705&  34.6176&$124.0\pm 40.2$&$  5.1\pm 37.1$\\
J1241+4617a  & 190.4097&  46.2881&$206.1\pm 37.2$&$...$\\
J1304+1202a  & 196.1090&  12.0460&$ 67.1\pm 32.3$&$122.1\pm 30.3$\\
J1353+5328a  & 208.3213&  53.4738&$ 99.7\pm 35.9$&$ 33.2\pm 33.1$\\
J1459+0024   & 224.7800&   0.4003&$278.1\pm134.8$&$...$\\
J1541+3153a  & 235.4727&  31.8915&$190.8\pm 24.4$&$ 77.3\pm 23.8$\\
Q0821+31     & 125.2817&  31.1264&$224.1\pm 24.9$&$ 85.0\pm 24.7$\\
Q0841+12a    & 131.1012&  12.7635&$381.2\pm 44.4$&$316.0\pm 43.6$\\
Q0930+28     & 143.4076&  28.7598&$ -3.3\pm 41.1$&$ 75.1\pm 37.6$\\
Q1209+09     & 182.8955&   9.0397&$...$&$237.7\pm 61.0$\\
Q1337+11     & 205.0102&  11.1082&$192.3\pm 50.9$&$...$\\
Q1502+48     & 225.6139&  48.6192&$240.4\pm108.6$&$...$\\
Q2223+20     & 336.4042&  20.6716&$378.5\pm 47.6$&$162.3\pm 43.4$\\
Q2342+34     & 356.2129&  34.5630&$326.1\pm 74.5$&$-20.2\pm 68.6$\\
SDSS0008-0958&   2.0639&  -9.9817&$199.3\pm 63.3$&$153.8\pm 57.0$\\
SDSS0013+1358&   3.3675&  13.9744&$486.0\pm 52.8$&$246.5\pm 50.7$\\
SDSS0016-0012&   4.0100&  -0.2069&$ 64.0\pm 33.4$&$ 95.6\pm 34.1$\\
SDSS0020+1534&   5.1207&  15.5766&$...$&$ 43.8\pm 87.2$\\
SDSS0035-0918&   8.7578&  -9.3049&$ 95.3\pm 35.6$&$ 36.6\pm 33.1$\\
SDSS0127+1405&  21.9492&  14.0953&$152.2\pm 58.4$&$...$\\
SDSS0128+1347&  22.1688&  13.7989&$ 68.9\pm101.8$&$139.8\pm 88.3$\\
SDSS0139-0824&  24.7558&  -8.4122&$262.7\pm 67.7$&$147.4\pm 65.3$\\
SDSS0142+0023&  25.5614&   0.3901&$228.9\pm 37.1$&$ 37.3\pm 36.7$\\
SDSS0225+0054&  36.4785&   0.9142&$231.1\pm125.1$&$129.3\pm109.4$\\
SDSS0234-0751&  38.5374&  -7.8521&$198.6\pm 68.0$&$266.2\pm 63.0$\\
SDSS0316+0040&  49.0410&   0.6786&$503.2\pm104.3$&$785.0\pm 90.7$\\
SDSS0751+3533& 117.9970&  35.5606&$294.1\pm124.4$&$123.4\pm107.6$\\
SDSS0826+3148& 126.5821&  31.8133&$377.3\pm 30.9$&$293.4\pm 30.7$\\
SDSS0828+4544& 127.1557&  45.7364&$-93.2\pm145.5$&$-55.1\pm128.9$\\
SDSS0840+4942& 130.1373&  49.7144&$193.1\pm177.7$&$ 21.4\pm156.7$\\
SDSS0902+5143& 135.5359&  51.7311&$-51.1\pm157.7$&$ 10.3\pm135.9$\\
SDSS0912+5621& 138.0959&  56.3578&$-42.8\pm 78.7$&$ 15.2\pm 70.1$\\
SDSS0912-0047& 138.1983&  -0.7881&$165.1\pm 68.3$&$ 19.8\pm 67.2$\\
SDSS0927+5621& 141.7746&  56.3539&$ 17.1\pm 81.2$&$123.6\pm 74.7$\\
SDSS0927+5823& 141.7870&  58.3887&$194.5\pm 52.5$&$ 87.5\pm 48.5$\\
SDSS1035+5440& 158.8092&  54.6778&$ 78.5\pm 74.2$&$ 73.8\pm 70.6$\\
SDSS1042+0117& 160.7180&   1.2933&$194.4\pm 76.3$&$ 44.6\pm 76.2$\\
SDSS1049-0110& 162.3143&  -1.1772&$108.1\pm 46.6$&$263.1\pm 43.5$\\
SDSS1131+6044& 172.8767&  60.7391&$ 54.8\pm 37.9$&$ 10.5\pm 37.7$\\
SDSS1151+0204& 177.8422&   2.0739&$-60.3\pm 88.6$&$-50.5\pm 86.6$\\
SDSS1155+0530& 178.9108&   5.5141&$171.9\pm 33.0$&$233.1\pm 30.8$\\
SDSS1208+6303& 182.0110&  63.0580&$139.5\pm 15.2$&$ 52.1\pm 14.8$\\
SDSS1235+0017& 188.9970&   0.2878&$ 40.1\pm151.3$&$122.2\pm140.7$\\
SDSS1249-0233& 192.3536&  -2.5608&$182.0\pm 37.3$&$175.7\pm 34.9$\\
SDSS1419+5923& 214.7763&  59.3867&$112.3\pm 36.7$&$ 52.1\pm 34.7$\\
SDSS1435+0420& 218.8039&   4.3433&$-86.5\pm178.0$&$234.0\pm170.5$\\
SDSS1610+4724& 242.5393&  47.4122&$111.8\pm 62.5$&$...$\\
SDSS1617+0028& 244.3243&   0.4742&$288.9\pm 87.7$&$-45.5\pm 80.4$\\
SDSS1658+3428& 254.5686&  34.4694&$156.3\pm 51.9$&$-32.6\pm 49.9$\\
SDSS1709+3258& 257.2887&  32.9676&$257.9\pm 57.9$&$222.1\pm 53.7$\\
SDSS2036-0553& 309.1762&  -5.8834&$342.4\pm 54.6$&$313.2\pm 50.9$\\
SDSS2044-0542& 311.1297&  -5.7110&$327.1\pm 73.3$&$115.9\pm 68.9$\\
SDSS2059-0529& 314.8434&  -5.4783&$214.0\pm 83.7$&$-75.8\pm 81.1$\\
SDSS2100-0641& 315.1043&  -6.6961&$333.1\pm 43.2$&$411.1\pm 41.4$\\
SDSS2141+1119& 325.3724&  11.3329&$584.5\pm 68.5$&$402.0\pm 63.7$\\
SDSS2151-0707& 327.8208&  -7.1315&$ 96.8\pm 49.2$&$ 93.8\pm 47.5$\\
SDSS2222-0946& 335.7338&  -9.7767&$170.9\pm 52.2$&$158.4\pm 49.3$\\
SDSS2231-0852& 337.9898&  -8.8701&$376.9\pm 71.7$&$ 12.9\pm 67.1$\\
SDSS2234+0057& 338.6605&   0.9583&$...$&$312.9\pm 30.7$\\
SDSS2238+0016& 339.6815&   0.2800&$294.9\pm 42.8$&$119.3\pm 40.0$\\
SDSS2244+1429& 341.2176&  14.4875&$299.2\pm 49.4$&$281.7\pm 47.6$\\
SDSS2315+1456& 348.9315&  14.9351&$160.9\pm 55.2$&$103.1\pm 55.4$\\
SDSS2343+1410& 355.9692&  14.1707&$501.3\pm 59.4$&$540.0\pm 55.0$\\
\tableline
\textbf{HIRES}\\
\tableline
Q2359          &   0.4583&  -1.9944&$...$&$157.8\pm 13.4$\\
J0108-0037     &  17.1117&  -0.6233&$...$&$ 87.5\pm 10.7$\\
Q0151+0448     &  28.4747&   5.0492&$268.0\pm 14.1$&$213.0\pm 13.8$\\
J022554+005451 &  36.4785&   0.9142&$219.7\pm 58.1$&$ 76.9\pm 45.7$\\
Q0551          &  88.1925& -36.6244&$...$&$122.1\pm 21.7$\\
HS0741+47T     & 116.3406&  47.5767&$...$&$284.8\pm 56.1$\\
FJ0812+32      & 123.1695&  32.1358&$232.8\pm 24.4$&$187.6\pm  7.9$\\
3C196          & 123.4000&  48.2175&$...$&$162.1\pm 71.6$\\
J0826+3148     & 126.5821&  31.8133&$319.9\pm 11.7$&$246.1\pm 15.1$\\
Q0841          & 131.1012&  12.7636&$...$&$303.7\pm 70.9$\\
J0900+4215     & 135.1396&  42.2628&$186.6\pm  7.1$&$104.4\pm  6.7$\\
J0929+2825a    & 142.3104&  28.4247&$111.4\pm  7.8$&$ 74.9\pm  9.9$\\
Q0930+28T      & 143.4076&  28.7597&$ 69.1\pm 20.1$&$ 60.6\pm 18.1$\\
J0953+5230     & 148.2879&  52.5083&$376.3\pm 19.7$&$ 86.3\pm 20.7$\\
J1010+0003     & 152.5758&   0.0642&$...$&$ 25.1\pm 31.5$\\
J1014+4300a    & 153.6966&  43.0083&$130.2\pm 10.6$&$ 80.0\pm  9.3$\\
J1035+5440a    & 158.8092&  54.6778&$120.2\pm 22.0$&$ 50.2\pm 14.0$\\
Q1104          & 166.6373& -18.3528&$184.4\pm  6.9$&$ 80.4\pm 13.1$\\
J113130+604420 & 172.8767&  60.7392&$-11.1\pm 21.1$&$ 15.6\pm 17.1$\\
HS1132+2243    & 173.7835&  22.4519&$  1.5\pm 23.0$&$ -9.5\pm 17.9$\\
J1155+0530     & 178.9108&   5.5142&$ 35.1\pm 29.7$&$...$\\
J1159+0112     & 179.9367&   1.2019&$177.2\pm 10.4$&$ 13.3\pm 10.5$\\
J1200+4015a    & 180.1659&  40.2656&$193.3\pm 16.7$&$114.9\pm 16.7$\\
J1211+0422     & 182.8233&   4.3728&$ 60.7\pm 16.7$&$ 56.7\pm 42.8$\\
Q1210+17       & 183.2628&  17.2397&$...$&$ 33.0\pm 11.6$\\
Q1215          & 184.3856&  33.0939&$ -0.2\pm 13.5$&$...$\\
J124020+145535a& 190.0871&  14.9267&$187.2\pm 37.6$&$ 81.9\pm 42.9$\\
J1304+1202a    & 196.1090&  12.0461&$192.5\pm 12.5$&$ 34.7\pm 15.1$\\
J1310+5424     & 197.6677&  54.4139&$174.8\pm 36.6$&$130.4\pm 13.4$\\
J1353+5328a    & 208.3213&  53.4739&$...$&$-22.5\pm 17.2$\\
PKS1354-17a    & 209.2753& -17.7339&$553.1\pm 74.8$&$203.2\pm 39.9$\\
J141030+511113a& 212.6275&  51.1872&$ 71.4\pm 33.1$&$ 41.3\pm 19.1$\\
J1435+5359     & 218.7521&  53.9983&$ 75.8\pm  7.8$&$ 21.3\pm  5.9$\\
J1541+3153a    & 235.4727&  31.8917&$233.0\pm 12.2$&$136.6\pm 15.7$\\
J1552+4910     & 238.1412&  49.1689&$136.3\pm 20.4$&$ 52.6\pm 16.8$\\
J1558-0031     & 239.5423&  -0.5222&$403.8\pm 15.9$&$300.1\pm 14.2$\\
J1604+3951a    & 241.0582&  39.8561&$134.1\pm 11.0$&$ 59.7\pm 14.6$\\
Q1759+75T      & 269.4433&  75.6544&$361.2\pm 14.5$&$223.9\pm 13.7$\\
Q2230          & 338.1471&   2.7986&$318.1\pm 25.9$&$240.6\pm 44.7$\\
Q2233+13       & 339.0800&  13.4389&$388.8\pm 30.6$&$311.5\pm 27.1$\\
FJ2334-09      & 353.6935&  -9.1367&$...$&$207.1\pm 25.9$\\
Q2343+12       & 356.6176&  12.8167&$299.6\pm  5.4$&$201.3\pm  7.4$\\
Q2344          & 356.6908&  12.7583&$198.6\pm 16.4$&$144.2\pm 18.8$\\
\enddata
\end{deluxetable}

\clearpage

\appendix

\section{Further Tests and Experiments}

In order to resolve the discrepancy between our measurement of the correlation and previous efforts, we have attempted a variety of 
changes, some of which are described below. Note that none of these changed our conclusions.

The \ebv\ values of MZ97 are based on fitting stellar models. We tried replacing them with corresponding values from SFD at the positions of their stars. As expected, since these are nearby stars and not all the way across the Galaxy, the SFD values are typically much higher, and the disagreement between our results and theirs becomes greater. In addition the scatter increases significantly indicating that the original values were far more appropriate. 

We have attempted correcting the SDSS spectra for the velocity of the ISM using \h1\ maps from \citet{hartmann97}. Every spectrum was de-redshifted by the median velocity we measured on its line of sight. This yielded no noticeable change, which is not surprising since the mean velocity in our fields is 11\,km\,s$^{-1}$ which amounts to sub \AA\ noise, and only a handful of targets are along directions with mean velocities greater than 50\,km\,s$^{-1}$. For simplicity we do not correct for these velocities throughout the paper. 

Another tracer of gas and dust is the \k1\ absorption line at $\lambda$7699. Repeating the procedure for the SDSS spectra around this wavelength we find that we cannot detect this line. This is consistent with the fact that while for \nad\ the weakest EWs we can measure are near 60\,m\AA, the EWs of \k1\ should be $\sim 30$\,m\AA\ according to MZ97.

We attempted replacing the SFD values with the previously often-used maps of \citet{burstein82}. While the resulting slope in EW vs \ebv\ is indeed shallower, it is only due to the increased noise in the measurements. Basically these maps are consistent with being a random reshuffling of the SFD maps with a dispersion of order 0.04\,mag. We similarly tried using the \citet{peek10} map, that refines the SFD map based on SDSS galaxy colors. We find a negligeable improvement in the agreement, but overall our departure from the MZ97 result is not resolved by their corrections. Again we do not use corrections from \citet{peek10} for the sake of simplicity. 

\subsection{A closer look at biases}\label{s:bias}

As discussed in Section \ref{s:compare}, the binning of the SDSS spectra introduces a bias, due to the extremely uneven population with a significant overabundance of low-extinction/low-EW spectra. The amount of bias depends on the prevalence of erroneous assignments to a given bin, which in turn depends on the precision of the SFD maps. Using simulations, we determined independently how much noise is present in the maps and how much bias need be corrected for in our measurements.

We picked a wavelength range to introduce a fake absorption line ($\lambda$5870), randomized a line width similar to those we measure for the \nad\ lines, assumed a range of correlation slopes, and varied the amount of noise we add to the \ebv\ values (we take a value of 100\,m\AA\ for the intrinsic dispersion in the correlation, as corroborated by our other tests). Regardless of the input slope, after we bin the spectra and apply the same procedure we used for the real data, the slope we recover is between 0.89 (no \ebv\ noise) and 0.65 ($\sigma_{E_{\rm (B-V)}}=0.1$) times the `real' slope. \citet{peek10} find that the maps they derive deviate from the SFD maps by less than 0.003\,mag typically, though they do find deviations of order 0.045\,mag is certain areas. Therefore the 12\% bias we obtain at $\sigma_{E_{\rm (B-V)}}=0.01$ seems the most appropriate. This is consistent with our finding in Section \ref{s:compare}, and our subsequent application of a 10\% correction.

\end{document}